\documentclass{PoS}
\usepackage{amsmath,amssymb}

\newcommand{\mathe}{\mathrm{e}}
\newcommand{\tmem}[1]{{\em #1\/}}
\newcommand{\tmop}[1]{\ensuremath{\operatorname{#1}}}
\newcommand{\tmtextbf}[1]{{\bfseries{#1}}}

\newcommand{\tmtexttt}[1]{{\ttfamily{#1}}}

\title{Strong coupling expansion Monte Carlo}

\ShortTitle{Strong coupling expansion Monte Carlo}

\author{\speaker{Ulli Wolff}\\
        Institut f\"ur Physik, Humboldt Universit\"at\\ 
        Newtonstr. 15 \\ 
        12489 Berlin, Germany\\
\\
        E-mail: \email{uwolff@physik.hu-berlin.de}}

\abstract{We give an overview on recently accomplished successful generalizations of
  `worm' or loop gas simulation methods to O($N$) and CP($N - 1$) sigma models
  and to simple fermion models. Beside the advantage of (practically)
  eliminated critical slowing down we also explain additional opportunities to
  estimate some observables with extremely improved signal to noise levels.
\begin{flushright} HU-EP-10/52 \end{flushright}
\begin{flushright} SFB/CCP-10-78 \end{flushright}
}

\FullConference{The XXVIII International Symposium on Lattice Field Theory, Lattice2010\\
		June 14-19, 2010\\
		Villasimius, Italy}
\begin{document}
\section{Introduction and conclusions}

In recent years a new approach to the simulation of statistical systems on the
lattice has been developed which goes under several names: world line or loop
gas formalism, all-order strong coupling (or hopping parameter) simulations or
`worm' algorithm methods. The key idea is to first reformulate the system as
the (complete) sum of its strong coupling graphs. This refers to a very simple
form of the strong coupling expansion that converges as long as the volume
remains finite. At large correlation length the graphs that have to be
included to obtain precise results are however of forbiddingly high order for
a systematic expansion. Using algorithms similar to those proposed in
{\cite{worm0}} and {\cite{prokofev2001wacci}} it has become possible on the
other hand to estimate the expansion of many observables by a Monte Carlo
procedure that samples a representative subset of contributions. An important
bonus is that this approach has been demonstrated to at least in some cases be
free of critical slowing down or free of sign problems where this is not so
with known methods in the conventional formulation. The problem of efficiently
producing independent {\tmem{long distance correlated}} field configurations
is translated into the need of efficiently passing between the relevant
{\tmem{large}} strong coupling graphs. This problem seems to be sufficiently
different to make progress in cases where for instance cluster algorithms in
the conventional setup do not work.

The topic has been reviewed before {\cite{Chandrasekharan:2008gp}} at Lattice
2008. It also seems to be closely watched by the finite-$\mu$ QCD community
and typically fills a subsection in their reviews, see
{\cite{deForcrand:2010ys}} and Sourendu Gupta's contribution to this
conference. These may be consulted in particular in connection with progress
on the sign problem. In the present contribution we mainly focus on the
important successful extension to non Abelian spin models of the O($N$) and
CP($N - 1$) type. While the method here is {\tmem{not}} confined to two
dimensions, most tests are conducted there because of asymptotic freedom and
the possibility to probe deeply into the continuum limit. We also cover
progress on fermions which unfortunately at present {\tmem{is}} confined to
two dimensions\footnote{See however
\cite{Chandrasekharan:2009wc} for ideas for an approximate method beyond $D=2$.
} 
and allows for instance simulations of the Gross Neveu model.
Finally we mention here that first steps have been made toward the treatment
of gauge theories, see {\cite{Tomz}}.

\section{The idea: Ising model as an example\label{secIsing}}

A good starting point to explain the strategy is the two point correlation in
the Ising model
\begin{equation}
  \langle \sigma (u) \sigma (v) \rangle = \frac{2^{- V} \sum_{\{\sigma (x) =
  \pm 1\}} \mathe^{\beta \sum_{l = \langle x y \rangle} \sigma (x) \sigma (y)}
  \sigma (u) \sigma (v)}{2^{- V} \sum_{\{\sigma (x) = \pm 1\}} \mathe^{\beta
  \sum_{l = \langle x y \rangle} \sigma (x) \sigma (y)}} = \frac{Z_2 (u,
  v)}{Z_0} . \label{I2point}
\end{equation}
Here $u, v$ are sites on a hypercubic periodic lattice of $V$ sites in
arbitrary dimension. In (\ref{I2point}) we emphasize the view of a correlation
as a ratio of two partition functions with and without field insertions.

In any {\tmem{finite volume}} the expansions{\footnote{An expansion in powers
of $\tanh \beta$ instead of $\beta$ would appear more efficient for the Ising
model, but would be less easy to generalize below.}} of $Z_0, Z_2$ in powers
of $\beta$ are convergent for all values of $\beta$. This includes the
vicinity of the critical point and all situations where Monte Carlo
simulations are performed. It will turn out that in general very high orders
in $\beta$ are required to realize this convergence and achieve precision.
This is not possible in a systematic truncated expansion as there are
unmanageably many terms or graphs. As in other physical cases it comes to
rescue that not all terms are needed. A Monte Carlo procedure will instead
sample a sufficiently `important' subset of high order terms. Usual systematic
strong coupling expansions are restricted to small correlation lengths but, on
the other hand, allow to take the thermodynamic limit of quantities like $Z_2
/ Z_0$ term by term in the expansion. It is through this step that a finite
radius of convergence emerges which in many cases (certainly in the Ising
model) corresponds to a physical phase transition.

\begin{figure}
\begin{center}
  \resizebox{6.5cm}{!}{\includegraphics{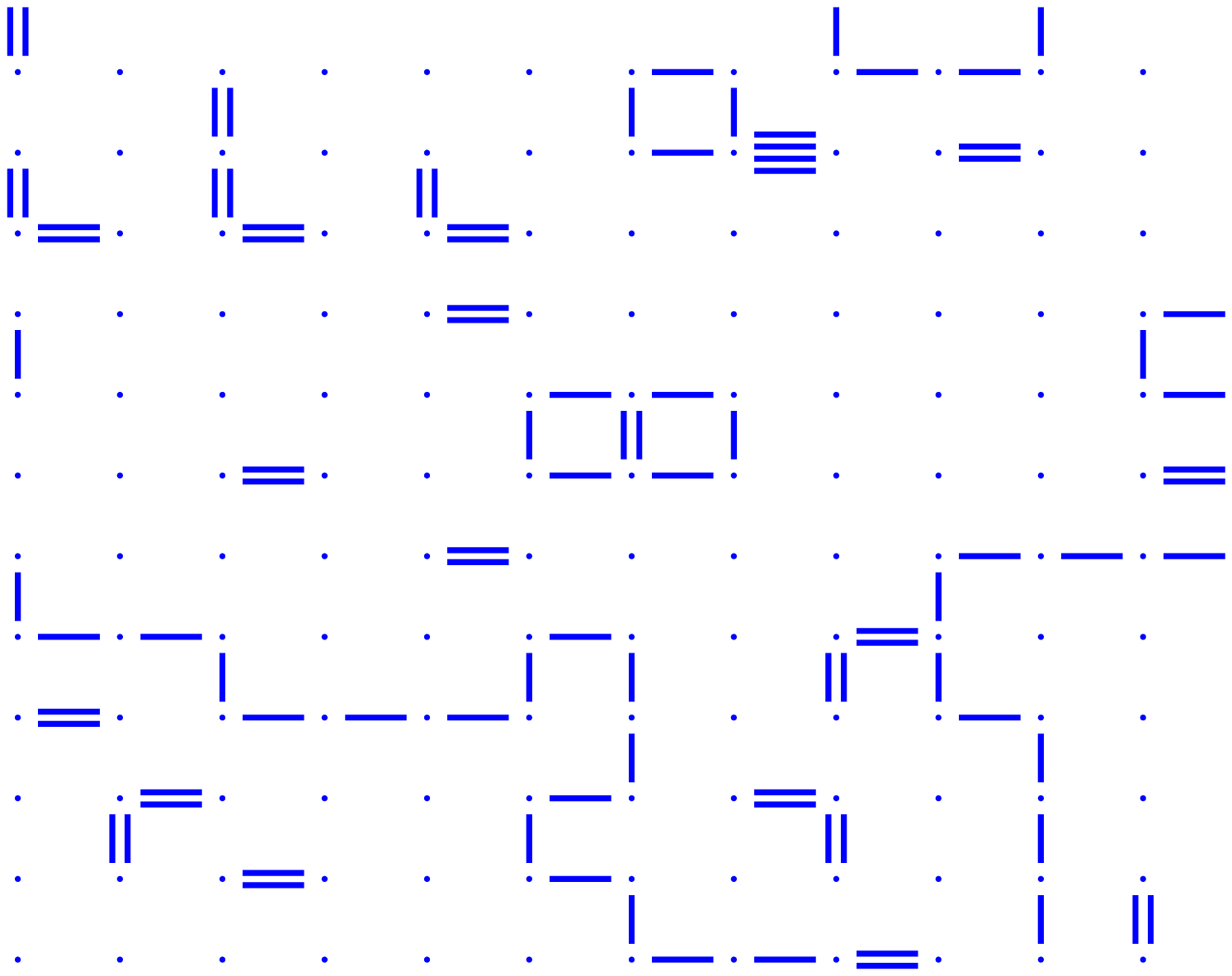}}
  {\hspace{2em}}\resizebox{6.5cm}{!}{\includegraphics{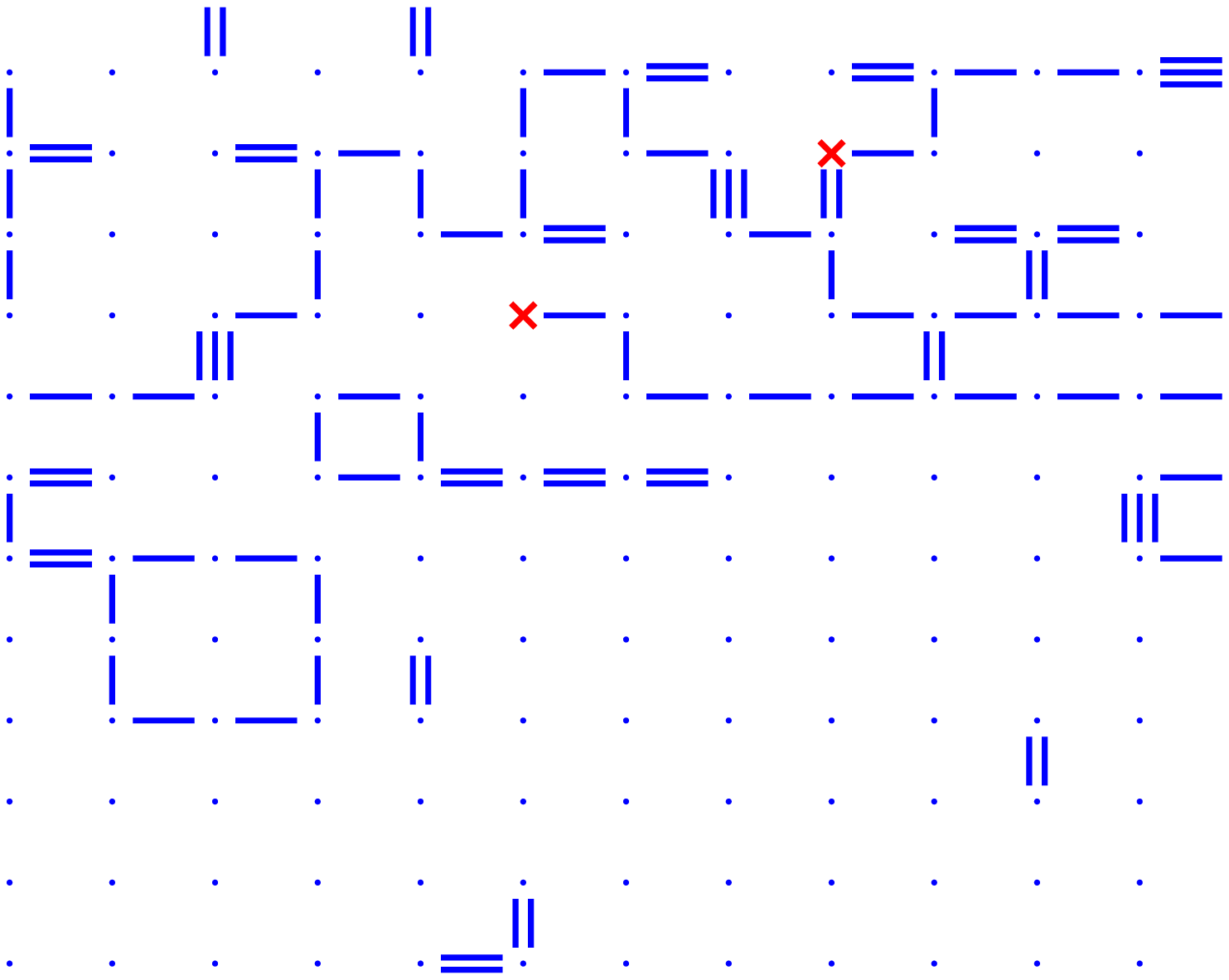}}
  \caption{A contribution in $\mathcal{G}_0$ (left graph) and $\mathcal{G}_2$
  (right graph). The boundaries are identified due to torus 
  boundary conditions\label{Igraphs}}
\end{center}
\end{figure}

The expansion is set up by using
\begin{equation}
  \mathe^{\beta \sigma (x) \sigma (y)} = \sum_{k = 0}^{\infty}
  \frac{\beta^k}{k!} \sigma (x)^k \sigma (y)^k \label{MrTaylor}
\end{equation}
for each neighbor pair on each link $l = \langle x y \rangle$ introducing
independent integers $k (l) = 0, \ldots, \infty$ on all links. For each
configuration $k \equiv \{k (l)\}$ the spins may now be summed over and the
partition functions are given as
\begin{equation}
  Z_0 = \sum_{g \in \mathcal{G}_0} \beta^{\sum_l k (l)} W [k], \hspace{1em}
  Z_2 (u, v) = \sum_{g \in \mathcal{G}_2 (u, v)} \beta^{\sum_l k (l)} W [k] .
  \label{Z02}
\end{equation}
In this formula the $k$-configurations are viewed as graphs of the type shown
in Fig. \ref{Igraphs}. There are $k (l)$ lines on each link. The left graph is
in the set $\mathcal{G}_0$ where the spin summation has enforced the constraint
that each site must be surrounded by an even number of lines and all different
such graphs make up $\mathcal{G}_0$.
The set $\mathcal{G}_2 (u, v)$ is visualized in the right graph in
Fig. \ref{Igraphs} with the red crosses at $u, v$ surrounded by an odd number
of lines due to the extra spin insertions. The factor
\begin{equation}
  W [k] = \prod_l \frac{1}{k (l) !}
\end{equation}
completes the weight implied by (\ref{MrTaylor}).

By differentiating $\tmop{lnZ}_0$ we derive the identity for a link $\langle x
y \rangle = l'$
\begin{equation}
  \beta \langle \sigma (x) \sigma (y) \rangle = \frac{1}{Z_0}  \sum_{g \in
  \mathcal{G}_0} \beta^{\sum_l k (l)} W [k] k (l') = \langle \langle k (l')
  \rangle \rangle_0 . \label{Ekeq}
\end{equation}
As indicated before a typical graph is thus O($V$) in $\beta$ close to the
critical point where the left hand side is O(1).

A direct simulation of the ensemble $Z_0$ in the form (\ref{Z02}) was tried a
long time ago in {\cite{Berg:1981jy}}. The authors designed a Monte Carlo
algorithm that samples graphs in $\mathcal{G}_0$ (mainly) by local
deformations over plaquettes. They observed critical slowing down comparable
to other standard methods. Probably mainly for this reason the approach does
not seem to have been pursued much further at the time. Also the accessibility
of physically interesting observables was not obvious in this formulation. The
two point function for example could in principle be estimated as a product
over strings of $k (l)$ but this would probably be inefficient at long
distance due to a large variance.

A breakthrough was achieved much later in {\cite{worm0}} and
{\cite{prokofev2001wacci}}. The essential idea was a joint simulation of $Z_0$
and $Z_2$ in an ensemble with the partition function
\begin{equation}
  \mathcal{Z}= \sum_{g \in \mathcal{G}_2} \beta^{\sum_l k (l)} W [k] =
  \sum_{u, v} Z_2 (u, v)
\end{equation}
where the sum over $\mathcal{G}_2$ without arguments is over graphs with all
possible insertion points
\begin{equation}
  \mathcal{G}_2 = \cup_{u, v} \mathcal{G}_2 (u, v) .
\end{equation}
Note that the graphs $\mathcal{G}_0$ contributing to $Z_0$ are also included
as diagonal contributions{\footnote{Each graph $g \in \mathcal{G}_0$ appear
$V$ times in the sum with all possible $u = v$. }} with $u = v$. Expectation
values are now defined as
\begin{equation}
  \langle \langle A \rangle \rangle = \frac{1}{\mathcal{Z}}  \sum_{g \in
  \mathcal{G}_2} \beta^{\sum_l k (l)} W [k] A [g] . \label{ZZO1}
\end{equation}
The identity (\ref{Ekeq}) now reads

\begin{equation}
  \beta \langle \sigma (x) \sigma (y) \rangle = \frac{\langle \langle k (l)
  \delta_{u, v} \rangle \rangle}{\langle \langle \delta_{u, v} \rangle
  \rangle} = \langle \langle k (l)  \rangle \rangle_0
  \label{Kdef}
\end{equation}
and summing over all links $l = \langle x y \rangle$ we measure
the internal energy of the original Ising model. In addition it is easy to
establish the connection for general correlations
\begin{equation}
  \langle \sigma (x) \sigma (0) \rangle = \frac{\langle \langle \delta_{x, u -
  v} \rangle \rangle}{\langle \langle \delta_{u, v} \rangle \rangle}
\end{equation}
which in particular implies $\langle \langle \delta_{u, v} \rangle \rangle =
\chi^{- 1} \geqslant V^{- 1}$ with the magnetic susceptibility $\chi$. The
fraction of sampled graphs that belongs to $\mathcal{G}_0$ gets smaller toward
the critical point but remains larger than one out of $V$.

In a very simple but useful generalization we include a nonnegative weight
$\rho^{- 1} (u - v)$, \ into our strong coupling ensemble
\begin{equation}
  \mathcal{Z}= \sum_{g \in \mathcal{G}_2} \beta^{\sum_l k (l)} W [k] \rho^{-
  1} (u - v) \hspace{1em} \Rightarrow \hspace{1em} \langle \sigma (x) \sigma
  (0) \rangle = \rho (x) \frac{\langle \langle \delta_{x, u - v} \rangle
  \rangle}{\langle \langle \delta_{u, v} \rangle \rangle} .
\end{equation}
We adopt the normalization $\rho (0) = 1$ and $\rho$ must respect the lattice
periodicity. The advantages of this modification will be discussed in sect.
\ref{secCPN}.

The essential move in a Monte Carlo simulation of the $\mathcal{G}_2$ ensemble
with partition function $\mathcal{Z}$ is now the following local update step.
One may move $u$ to one of its nearest neighbors by shifting it over one of
the $2 D$ links attached to it. At the same time the $k (l)$ of that link is
changed by $\pm 1$ (adding or removing a line of $g$). Of course similar moves
may be made at $v$, alternatingly or picking randomly one of the two insertion
points. These allowed moves staying within $\mathcal{G}_2$ may be used now as
proposals for Metropolis acceptance steps. We do not describe any realization
here in all details, but refer to the literature. Concrete procedures for the
$\beta$ expansion discussed here may be found in {\cite{prokofev2001wacci}} or
{\cite{Wolff:2009ke}}. Algorithms for the $\tanh \beta$ expansion of the Ising
model with $k (l) \in \{0, 1\}$ are discussed in detail in
{\cite{Wolff:2008km}}, {\cite{Deng:2007jq}}.

In these papers it is numerically demonstrated that strong coupling
simulations of the Ising model have very much reduced and in many cases
completely eliminated critical slowing. We conclude that it is advantageous to
enlarge the graph space from $\mathcal{G}_0$ to $\mathcal{G}_2$ by allowing
defects. This is true even if we measure in $\mathcal{G}_0$ only as in
(\ref{Kdef}), but we have seen that the `intermediate' graphs contain even
more interesting information. In these proceedings we consider the Ising
discussion only as a preparation for more elaborate models and hence do not
review performance results here in more detail.

\section{ Nonlinear sigma models}

\subsection{O($N$)}

In {\cite{Wolff:2009kp}} a generalization of the above strong coupling graph
representation to the O($N$) invariant nonlinear sigma model has been given.
In this case $Z_2$ generalizes to
\begin{equation}
  Z_2 (u, v) = \left[ \prod_z \int_{} d^N s \delta (s^2 - 1) \right]
  \mathe^{\beta \sum_{l=\langle xy \rangle} s (x) \cdot s (y)} s (u) \cdot s (v) \label{Zuv}
\end{equation}
where $s (x)$ is an $N$ component unit vector integrated over the sphere. For
$N = 1$ the Ising model is obviously recovered. To generate an expansion in
$\beta$ we again expand the Boltzmann factor on each link. To then integrate
out the spins for each term in this expansion we need the integral over the
sphere with an arbitrary monomial in the spin as integrand. This information
follows by differentiation of the generating function with an $N$ component
source $j$
\begin{equation}
  \int_{} d^N s \delta (s^2 - 1) \mathe^{j \cdot s} = \sum_{n = 0}^{\infty} c
  [n ; N] (j \cdot j)^n
\end{equation}
with coefficients
\begin{equation}
  c [n ; N] = \frac{\Gamma (N / 2)}{2^{2 n} n! \Gamma (N / 2 + n)}
\end{equation}
deriving from the expansion of modified Bessel functions. Working out the
combinatorics, i.e. the multiplicities of each term, we arrive
at{\footnote{The symbols $\mathcal{G}_2, W$ and later  $\mathcal{S}$ are re-used for the
different classes of models that we discuss although their precise form
becomes context dependent in this way. Factors included in $W$ in the
references are sometimes pulled out and made explicit in this write-up. }}
\begin{equation}
  \mathcal{Z}= \sum_{g \in \mathcal{G}_2} \beta^{\sum_l k (l)} W [k ; N]
  \frac{N^{|g|}}{\mathcal{S}[g]} \times \rho^{- 1} (u - v) . \label{ZZON}
\end{equation}
\begin{figure}\centering
  \resizebox{!}{4cm}{\includegraphics{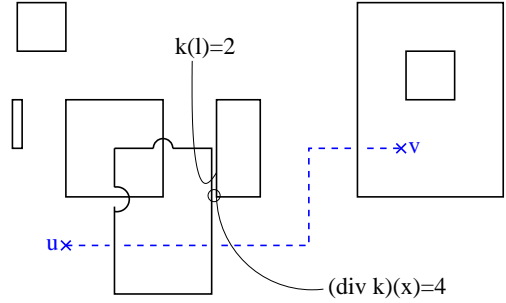}}
  \caption{Schematic view of a graph $g \in \mathcal{G}_2$ for the O($N$)
  model. The divergence div $k$ means the number of lines around a
  site.\label{ONgraph}}
\end{figure}

{\noindent}Several elements in this formula require explanation:
\begin{itemize}
  \item The graphs $\mathcal{G}_2$ now differ from those in Fig. \ref{Igraphs}
  only in so far that the even number of lines surrounding a site are
  connected pair-wise except two lines ending at $u, v$. This is visualized
  for a simple case in Fig. \ref{ONgraph}.
  
  \item By the pairings (corresponding to contractions of the $N$ spin
  components) a number of $|g|$ closed loops form in the graph $g$, each
  contributing a factor $N$.
  
  \item The weight $W$ collects $k (l) !$ factors and the $c [n ; N]$ from the
  site integrations.
  
  \item The symmetry factor $\mathcal{S}[g]$ generically equals unity. Only if
  a graph has extra symmetries under the exchange of lines then it equals the
  order of this symmetry group. This is analogous to symmetry factors in
  Feynman diagrams. More details on this subtlety are found in the erratum of
  {\cite{Wolff:2009kp}}.
\end{itemize}

The relation between the spin correlation and the graph ensemble is changed
only by the spin contraction
\begin{equation}
  \langle s (x) \cdot s (0) \rangle = \rho (x) \frac{\langle \langle
  \delta_{x, u - v} \rangle \rangle}{\langle \langle \delta_{u, v} \rangle
  \rangle} .
\end{equation}
The simulation of (\ref{ZZON}) requires in addition to the update steps
outlined before reroute moves where the local line connectivity is changed.
It suffices to go to $u$ (or $v$) and to randomly divorce one of the
line-pairs passing through\footnote{
If there are no such lines (as in Fig.~\ref{ONgraph}), no move is made.
}. The single line previously ending at $u$ is
remarried to one of the divorcees with the other one becoming the new single
line. This is again employed as a Metropolis proposal that is accepted with a
well defined probability dictated by the weights in (\ref{ZZON}).
By taking into account the asymmetric a priori proposal probabilities,
also $\mathcal{S}[g]$ is seen to be implemented correctly.

To implement the simulation just sketched, the graph structure including its
connectivity has to be mapped on a discrete structure in the Computer. This is
possible by a linked list. In the language C the configuration can be coded
into pointers where each line element residing on a link gets a name and
points to its successor and predecessor along its closed loop. In addition
there must be integer variables describing the geometrical embedding of the
graph on the lattice.

The weight $N^{|g|}$ may be implemented either exactly or stochastically. In
the original publication {\cite{Wolff:2009kp}} the exact algorithm
(R-algorithm) has been described. In this case the $N$-dependent weight
(\ref{ZZON}) is fed into the Metropolis decision. Then it must be known in the
reroute step, if the passing through line, that is picked for swapping the
connectivity, belongs to the line connecting $u$ and $v$ or if it is part of a
separate closed loop. To make this nonlocal information available requires to
sometimes travel around one of the closed loops by following the corresponding
pointers. As typical loop circumferences grow in the continuum limit, an
elementary reroute step costs more than O(1) operations. The numerical observation
in {\cite{Wolff:2009kp}} can now be summarized as follows. There is
(practically at least) no slowing down in units of iterations corresponding to
order O($V$) elementary steps. As they cost however slightly more than O($V$)
operations there is a small effective critical exponent. It was estimated
around $z \approx 0.3$. This refers to the O(3) model in $D = 2$ with large
volume and correlation lengths $\xi = 7, \ldots, 65$ and to the critical $D =
3$ model at $L = 32, 64$. The name R-algorithm derives from the fact that here
$N$ may be taken also to non-integer values by continuing the weights. For integer
$N$ one may formulate the I-algorithm where the weight $N^{|g|}$ is
incorporated stochastically. Then each closed loop as well as the line between
$u$ and $v$ carries an integer degree of freedom $i = 1, 2, \ldots, N$ that is
independently summed over. In the reroute step, only lines with the same $i$
can join. Additional update steps are now needed to move the $i$-labels. A
minimal way to do this is to randomly assign a new label to the line between
$u$ and $v$ after O($V$) elementary steps. In a short test Tomasz Korzec has
verified that this form of the I-algorithm shows very little critical slowing down
in the O(3) model for $D = 2$ and $\xi$ in the range mentioned
before. A more detailed description of the R- versus I-algorithm together with
numerical results is given in
the paper {\cite{Wolff:2010qz}} about the loop formulation of the CP($N - 1$)
model to which we come in sect. \ref{secCPN}.

\subsection{O(3) model with Nienhuis action}

We now imagine to restrict the graph summation in (\ref{ZZO1}) or (\ref{ZZON})
to the subclass of graphs which obey the constraint $k (l) \leqslant 1$ on all
links, which can be easily implemented in the simulations. In the Ising model
this changes the $\beta$ expansion into the tanh$\beta$ expansion. Thus, if we
accompany the restriction by this substitution we obtain exactly the same
correlation and hence $u - v$ distribution as before. In general the reduced
set of graphs is equivalent to starting from
\begin{equation}
  Z_2 (u, v) = \left[ \prod_z \int_{} d^N s \delta (s^2 - 1) \right] \left[
  \prod_{l = \langle x y \rangle} \left\{ 1 + \tilde{\beta} s (x) \cdot s (y)
  \right\} \right] s (u) \cdot s (v) \label{ZuvN}
\end{equation}
instead of (\ref{Zuv}). This is clearly a new lattice model -- hence we rename
$\beta \rightarrow \tilde{\beta}$ -- and the question arises if it belongs to
the same universality class despite its strange appearance. Such an action has
been introduced before {\cite{Domany:1981fg}} and studied in detail by
Nienhuis {\cite{Nienhuis:1982fx}}. An exact solution was obtained by summing
the strong coupling expansion on honeycomb lattices where the loops now cannot
intersect. The critical region could be reached for $N \leqslant 2$ with
$\tilde{\beta} \leqslant 1$, i.e. a nonnegative weight in (\ref{ZuvN}) and
universality was supported.
\begin{figure}\centering
  \resizebox{!}{7cm}{\includegraphics{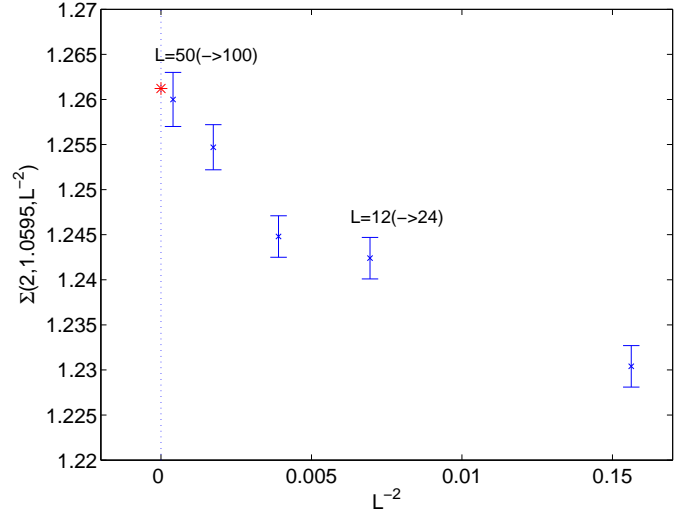}}
  \caption{Continuum extrapolation of a step scaling function based on
  (\protect\ref{ZuvN}).
\label{ssf}}
\end{figure}
In our simulations there is absolutely no sign problem when taking
$\tilde{\beta} > 1$ where in the original path integral there seems to be a
drastic sign problem. We thus simulated the O(3) model with $k (l) \leqslant 1$ and
quickly found that no criticality was reached with $\tilde{\beta} \leqslant
1$. To investigate universality we computed a step scaling function
{\cite{Luscher:1991wu}} for the finite volume mass gap extracted from time
slice correlations. In Fig. \ref{ssf} we see that the data points accurately
extrapolate to the star which is the exact universal answer known
{\cite{Balog:2003yr}} for this case. These runs involve values of
$\tilde{\beta}$ in the range $1.8 \ldots 3.1$. Our careful conclusion is that
at least for this special case a universal result is reproduced by
(\ref{ZuvN}) at a significant precision.

\subsection{CP($N-1$) \label{secCPN}}

Another class of nonlinear sigma models that are of physical interest are the
$\tmop{CP} ( N - 1)$ systems. There the spins label one dimensional subspaces
in complex space and may be parameterized by $\phi (x) \in \mathbb{C}^N, |
\phi (x) | = 1$ where $\phi$ differing by a phase have to be identified.
There are two popular lattice actions compatible with this structure. One is
the explicit gauge field action
\begin{equation}
  - S [\phi, U] = \beta \sum_{x \mu} [U (x, \mu) \phi^{\dag} (x) \phi (x +
  \hat{\mu}) + U^{- 1} (x, \mu) \phi^{\dag} (x + \hat{\mu}) \phi (x)]
  \label{SU}
\end{equation}
where nearest neighbors are coupled with a U(1) gauge field $U (x, \mu)$. It
is independently integrated over without an action of its own. As it can
absorb local phase changes to $\phi$ the geometric structure of the model
is respected. A second option is provided by the quartic action
\begin{equation}
  - S_q [\phi] = 2 \beta_q \sum_{x \mu} | \phi^{\dag} (x) \phi (x + \hat{\mu})
  |^2 \label{Sq}
\end{equation}
which exhibits local U(1) invariance without extra fields. The standard
expectation is that these actions are in the same universality class and
produce the same continuum quantum field theory.

A convenient way to probe the model is by correlations of the adjoint local
density \ $j^a (x) = \phi^{\dag} (x) \lambda^a \phi (x)$, where
$\lambda_a$ are a basis of hermitian traceless $N \times N$ matrices
normalized by $\tmop{tr} (\lambda^a \lambda^b) = 2 \delta^{a b}$, i.e.
generalized Gell-Mann matrices. Again we focus on the two point function
\begin{equation}
  \langle j^a (u) j^a (v) \rangle = \frac{Z_2 (u, v)}{Z_0} .
\end{equation}
By steps that generalize those of the Ising and O($N$) cases and which can be
found in detail in {\cite{Wolff:2010qz}} we construct a strong coupling
representation (first for (\ref{SU}))
\begin{equation}
  \mathcal{Z}= \sum_{g \in \mathcal{G}_2} \beta^{2 \sum_l k (l)} W [k ; N]
  \frac{N^{|g|}}{\mathcal{S}[g]} \times \rho^{- 1} (u - v) . \label{ZZCPN}
\end{equation}
The complex field variables lead to a modified graph structure
$\mathcal{G}_2$:
\begin{itemize}
  \item Each line on a link carries an orientation (arrow) and they are paired
  at the sites in a way respecting the sense of arrows.
  
  \item There are two lines of opposite orientation running between $u$ and
  $v$.
  
  \item On each link there is the same number of arrows in either direction.
\end{itemize}
The last constraint is a direct consequence of integrating out the U(1) gauge
field. In the exponent $k (l)$ is the number of lines {\tmem{per
orientation}}. The weight $W$ is again a local product of explicitly known
{\cite{Wolff:2010qz}} terms and $\mathcal{S}[g]$ is the symmetry factor. The
connection with the adjoint correlation can in this case be written as
\begin{equation}
  \langle j^a (0) j^b (x) \rangle = \rho (x) \frac{2 \delta_{a b}}{N (N + 1)} 
  \frac{\langle \langle \delta_{u - v, x} \rangle \rangle}{\langle \langle
  \delta_{u, v} \rangle \rangle} . \label{corr}
\end{equation}
If we repeat the construction starting from the quartic action we arrive at
exactly the same graph structure but obtain a different expression for $W$ and
have to replace $\beta^2 \rightarrow \beta_q$.

\begin{figure}\centering
  \resizebox{!}{7cm}{\includegraphics{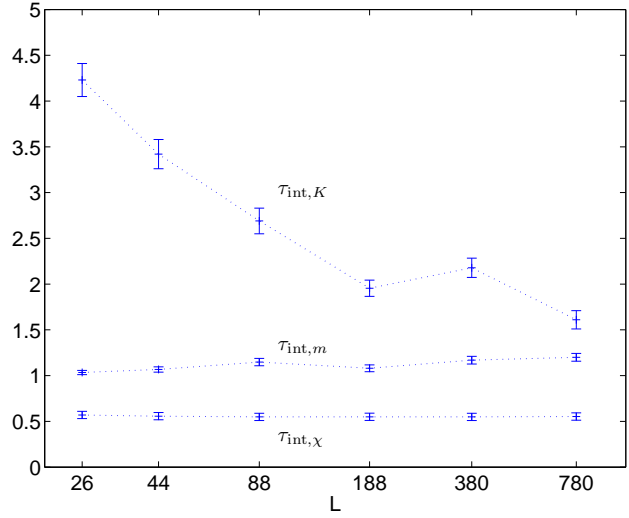}}
  \caption{Autocorrelation times in the two dimensional CP(3) model in Monte
  Carlo time units comparable to `sweeps'.\label{taus}}
\end{figure}

It turns out {\cite{Wolff:2010qz}} that this model may be simulated by a
procedure very similar to the one described for the O($N$) model. For the
I-implementation treating the factor $N^{|g|}$ stochastically we show measured
integrated autocorrelation times in Fig. \ref{taus}. We consider a series of
simulations based on (\ref{SU}) with $D = 2, N = 4$ and fixing $L / \xi
\approx 10$. The observable $K$ is defined as the right hand side of
(\ref{Kdef}) (summed over $l$) which also here is equivalent to the internal
energy. The susceptibility $\chi$ and the mass $m$ are extracted via the
correlation (\ref{corr}).

By computing a step scaling function the universality between the two lattice
realizations of the CP($N - 1$) model with actions (\ref{SU}) and (\ref{Sq})
has been confirmed at high precision (see Fig. 4 in {\cite{Wolff:2010qz}}) for
$D = 2, N = 3$.

In CP($N - 1$) models topology is of special interest. The status here is
that the extension of (\ref{ZZCPN}) to include a $\theta$ term is given in
{\cite{Wolff:2010qz}}. A simulation of this modified system remains to be
tested. Amplitudes in this case are not strictly positive any more, but it is
not known for which values of $\theta$ this leads to numerically problematic sign
fluctuations.

\begin{figure}\centering
  \resizebox{!}{10cm}{\includegraphics{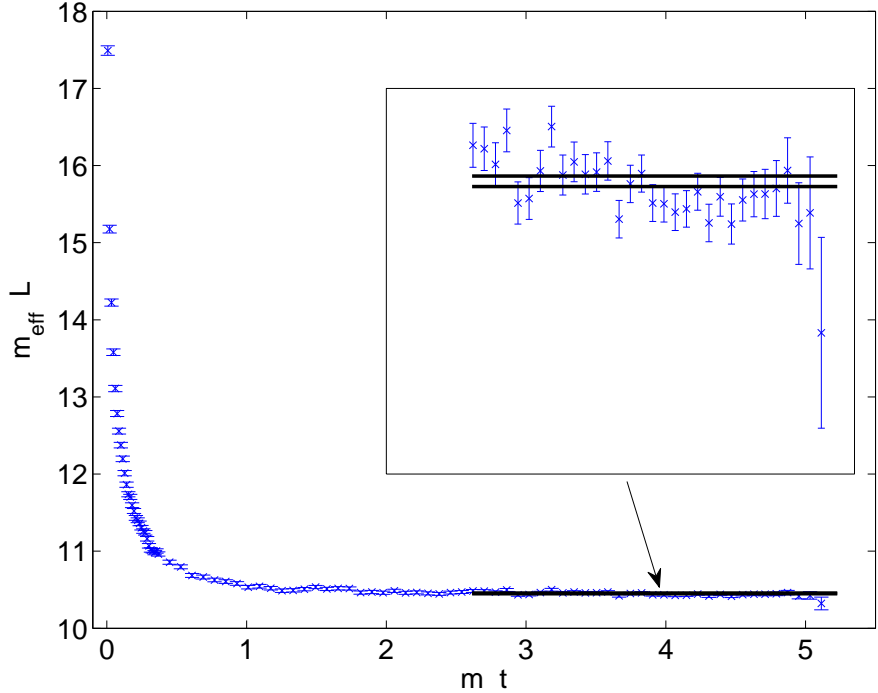}}
  \caption{Effective mass as a function of time slice separation $t$ for $L =
  780 \approx 10 \xi$ in the CP(3) model.\label{meff}}
\end{figure}

It is now time to come back to the usage of the free weight $\rho$ in our
simulations. We read (\ref{corr}) as follows: If we are able to guess the
behavior of the two point function and use this guess for $\rho (x)$, then, up
to known factors, the histogram $\langle \langle \delta_{u - v, x} \rangle
\rangle$ yields the correction factor that turns our guess into the exact
answer. For a perfect guess, it would be constant, in other words all possible
separations $u - v$ would occur with equal frequency. Up to autocorrelation
effects we would then expect equal relative statistical errors {\tmem{at all
separations}}. One can thus trace the exponential decay without degrading
signal to noise ratio. That this really works in practice is shown in Fig.
\ref{meff}. Here $\rho$ was set to the scalar propagator with mass $10 / L$.
Similar plots are available for Ising and O($N$) simulations.

\section{Fermions}

\begin{figure}\centering
  \resizebox{6cm}{!}{\includegraphics{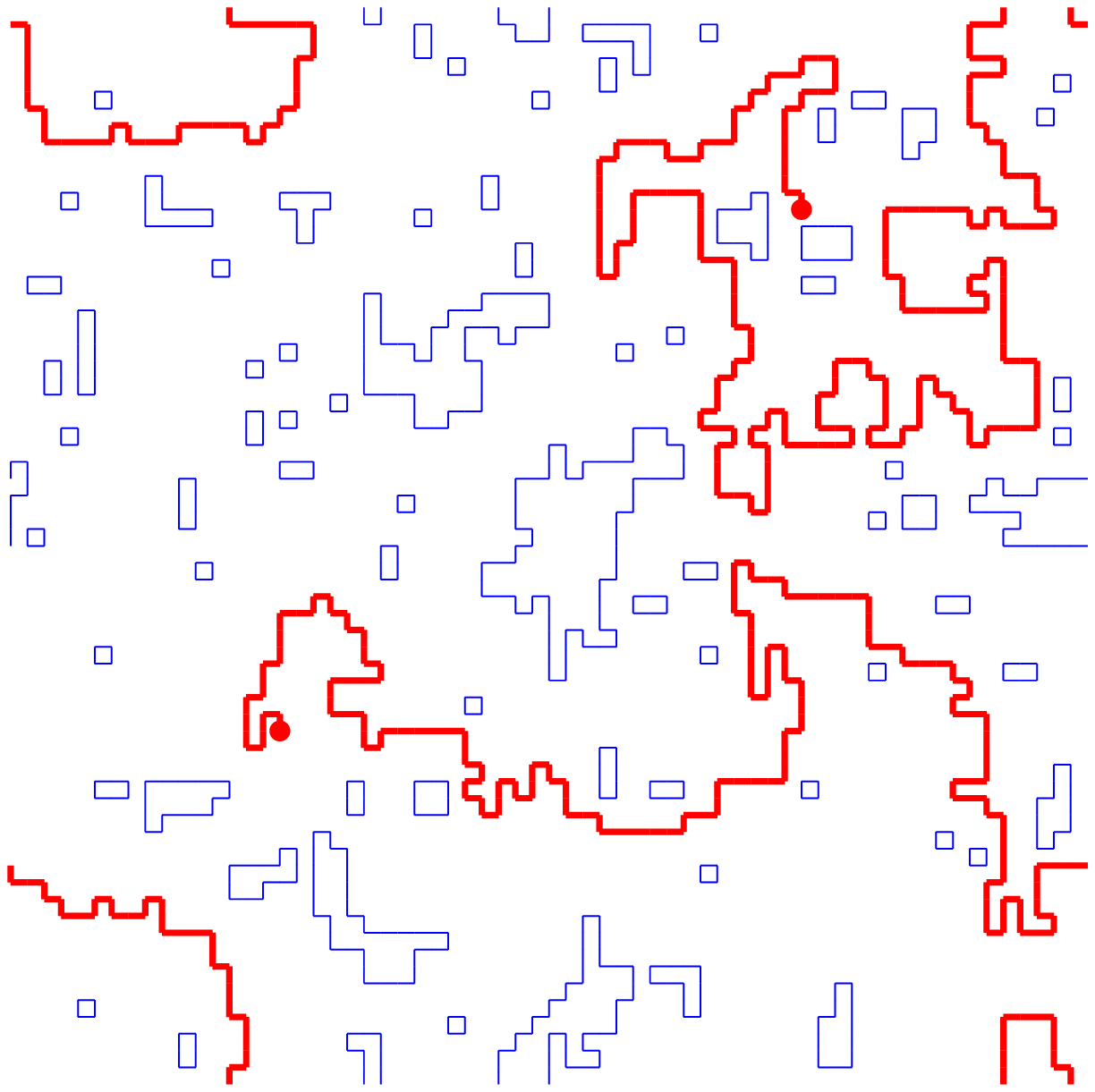}} \ \
  \resizebox{8cm}{!}{\includegraphics{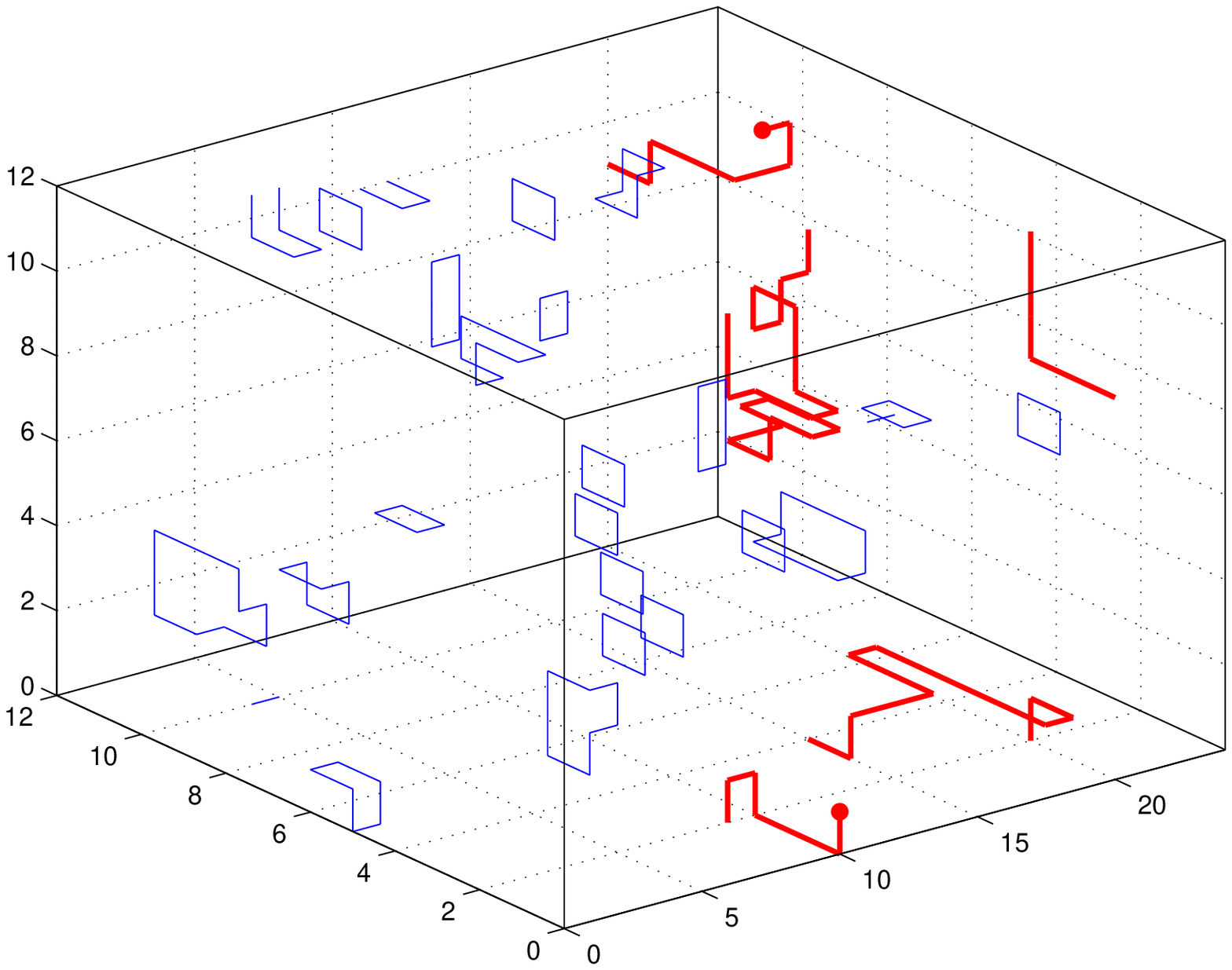}}
  \caption{Loop or hopping parameter expansion graph configuration for a
  Majorana fermion in $D = 2$ (left panel) and $D = 3$ (right
  panel).\label{fermfig}}
\end{figure}

In {\cite{Wolff:2007ip}} the problem of a 2D Majorana-Wilson fermion in an
external scalar field on a two dimensional torus was formulated as a loop gas
and simulated by cluster methods. Urs Wenger first proposed
{\cite{WengerLeilat08}} that in this formulation this system could
alternatively be simulated with the `worm' algorithm
{\cite{prokofev2001wacci}} which then triggered the independent study
{\cite{Wolff:2008xa}}. In this case the relevant partition function with two
insertions is
\begin{equation}
  \langle \xi_{\alpha} (u) \overline{\xi}_{\beta} (v) \rangle = \left[ \prod_z
  \int d \xi_1 d \xi_2 \mathe^{- \frac{2 + m}{2} \overline{\xi} \xi} \right]
  \left[ \prod_{l = \langle x y \rangle} \mathe^{\overline{\xi} (x) P (
  \widehat{y - x}) \xi (y)} \right] \xi_{\alpha} (u) \overline{\xi}_{\beta}
  (v) .
\end{equation}
We here integrate over two Grassmann variables per site, $P ( \widehat{y -
x})$ is the Wilson projector $\frac{1}{2} (1 - n_{\mu} \gamma_{\mu})$ if the
link is $y = x + n$ and $\overline{\xi}$ stands for $\xi^{\top} \mathcal{C}$
with charge conjugation $\mathcal{C} \gamma_{\mu} \mathcal{C}^{- 1} = -
\gamma_{\mu}^{\top}$. An external field is present if $m = m (x)$ is not
constant. Due to the projector nature of $P$ and the Grassmann nilpotency, the
expansions of the link factors have only two terms each ($k (l) = 0, 1$) as
for the Nienhuis action. Moreover there can be at most two lines adjacent to a
site and as a consequence loops and the line between $u$ and $v$ cannot
intersect. A typical configuration is shown in the left panel of Fig.
\ref{fermfig} for a massless free fermion.
In this case the correlation is given by
\begin{equation}
  \langle \xi_{\alpha} (x) \overline{\xi}_{\beta} (0) \rangle = \rho (x)
  \frac{\langle \langle \delta_{x, u - v} \Phi (k) M_{\alpha \beta} (k)
  \rangle \rangle}{\langle \langle \delta_{u, v} \Phi (k) \rangle \rangle}
\end{equation}
where $\Phi (k)$ is a well defined and readily computable sign and $M_{\alpha
\beta} (k)$ is a set of $2 \times 2$ matrices which only depends on the
directions in which $u$ and $v$ are approached by their connecting line.
Although the observable can change sign, this two dimensional fermion has no
serious sign problem, as neither the denominator nor the numerator gets very
small. This is because the bulk of closed loops that close without winding
around the torus are all positive. This is a specialty of two dimensional
fermions which here appears as the following feature. For each closed loop
there is the usual fermionic minus sign, but it is canceled by another sign.
The latter arises from the trace of a string of Wilson projectors multiplied
along the loops. This minus sign can be understood as the sign that arises
when a spinor is parallel transported around the loop and is rotated by $2
\pi$. For subtleties related to (anti)periodic boundary conditions of the
finite system the reader is referred to {\cite{Wolff:2008xa}} and to
{\cite{Bar:2009yq}}. In this publication it is also shown how $N$ species of
this loop gas can be coupled to represent and simulate the O($N$) invariant
Gross Neveu model in a remarkably efficient way.

The loop representation of the Majorana fermion was generalized from $D = 2$
to $D = 3$ in {\cite{Wolff:2008xa}}. Here we still have only two Dirac
components and the efficient computability is unchanged. In the right panel of
Fig. \ref{fermfig} we see a configuration from a simulation of a free fermion
at $m = 0.65$ on a $12^2 \times 24$ lattice where we reproduced the propagator
very precisely. Making however the mass smaller and/or the system much larger
we very abruptly encounter the full-blown sign problem. The reason is that in
$D = 3$ the positivity of loops only holds as long as they are planar. If non
planar loops become abundant, the spin factor assumes all values in Z(8) and
does not cancel the Fermi minus any longer. Instead of Z(8) there would
formally be a U(1) in the continuum, which is reduced to Z(8) on the lattice
along with rotations being reduced to the hypercubic subgroup. This fermionic
sign problem is presently unsolved. We consider the simple $D = 3$ Majorana
fermion as a good laboratory for further thinking.

\section{Triviality of $\varphi^4$ theory}

In four dimensions the number one textbook example for a quantum field theory
with a self-interacting real scalar field is believed to be trivial, i.e. a
free field, once the continuum limit is taken. While this is rigorously known
to be the case for $D > 4$ and false for $D < 4$ in the borderline case of $D
= 4$ the belief in triviality rests on numerical demonstrations. As a
byproduct of the strong coupling reformulation discussed here, we have found a
very much improved handle on such numerical checks for the Ising limit of
$\varphi^4$ which is the most interesting parameter range for triviality. One
of the techniques to obtain rigorous bounds in $D > 4$ has been developed by
Aizenman {\cite{Aizenman:1982ze}}. He uses nothing but the all-order strong
coupling form that we have developed in sect. \ref{secIsing}, \ called random
current representation by him. Quantized currents $k (l)$ flow through the
links and are conserved mod 2 at sites, with two sources at $u$ and $v$. By
borrowing his replica trick and a graph theoretical proposition we could
establish {\cite{Wolff:2009ke}} the following identity for the Ising model at
arbitrary $D$ and volume $L^D$
\begin{equation}
  g_R = - \frac{\chi_{_4}}{\chi^2} (m_R^{})^D = 2 z^D \langle \langle
  \mathcal{X} \rangle \rangle_{(g, g') \in \mathcal{G}_2 \times \mathcal{G}_2}
  . \label{gR}
\end{equation}
In this formula for the usual renormalized coupling $g_R$
\begin{itemize}
  \item $\chi$ is the 2-point susceptibility,
  
  \item $\chi_4$ is the 4-point (connected, symmetric phase) susceptibility,
  
  \item $m_R$ is a renormalized mass, for example using the second moment
  definition,
  
  \item any fixed values of $z = m_R L$ defines a renormalization scheme,
  
  \item the simulation samples two independent replica of graphs $g, g'$ with
  corresponding\\ $k, k', u, u', v, v'$,
  
  \item $\mathcal{X} \in \{0, 1\}$ is an observable computed as follows.
  $\mathcal{X}= 1$ holds iff all four defects are in {\tmem{one cluster}} of
  an auxiliary bond percolation problem. The bond variables in this problem
  are `off' only on links where $k (l) = 0 = k' (l)$ holds and `on' on all
  others. From cluster simulations we know how to efficiently compute
  $\mathcal{X}$. 
\end{itemize}
Note that (\ref{gR}) implies $\chi_{_4} \leqslant 0$, i.e. the Lebowitz
inequality is manifest in this estimator. The advantage of our (Aizenman's)
method lies in not having to perform a numerical cancellation to compute
$\chi_{_4}$ which avoids a large significance loss. Triviality now amounts to
the question whether or not $g_R \searrow 0$ as $L / a \rightarrow \infty$.
Here for each $L / a$, $\beta$ is determined by tuning $z$ to the chosen value.

\begin{figure}\centering
  \resizebox{10cm}{!}{\includegraphics{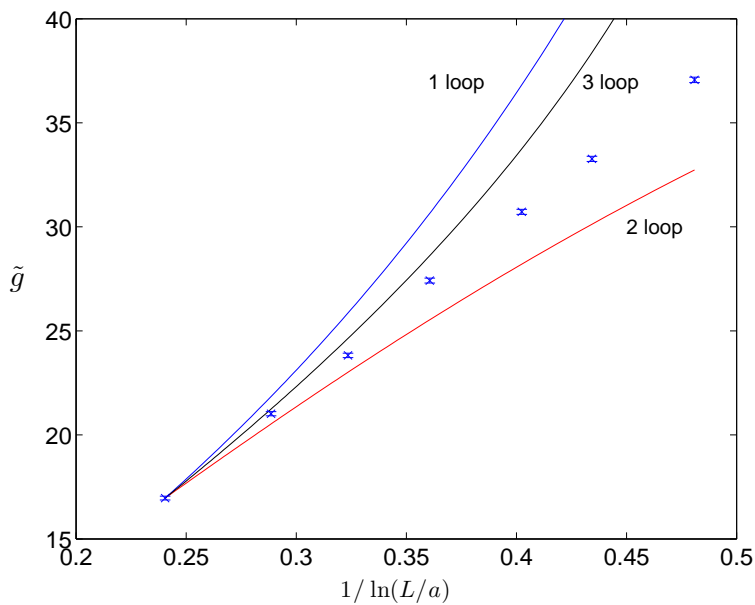}}
  \caption{Cutoff dependence of the renormalized coupling for $D = 4$, $z =
  4$. By starting integrations of the Callan Symanzik equation at the
  leftmost data point the lines are produced.\label{trivz4}}
\end{figure}

In {\cite{Wolff:2009ke}} a study was made for a relatively small volume $z =
2$. Although the results for $D = 3, 4, 5$ are consistent with the triviality
expectations, in $D = 4$ there remained some tension in matching with the
perturbative coupling evolution close to the continuum limit. Probably this
must be attributed to the weakly damped fluctuations of the constant mode. We
therefore made another study with $z = 4$ shown in Fig. \ref{trivz4}.
The data points with $L / a = 8, \ldots, 64$ have errors of only
about the symbol size in spite of only modest CPU time invested on some PCs.
The coupling $\tilde{g}$ may be identified with $g_R$ here. We see a
convincing `convergence' of the perturbative evolution toward the data points at
small $a / L$ (left side in the plot). If we are willing to conclude that this
agreement persists for yet smaller $a / L$ then triviality is established for
$\varphi^4$ (in the Ising limit).

\acknowledgments
I would like to thank Tomasz Korzec for discussions and numerical checks.
Financial support of the DFG via SFB transregio 9 is acknowledged.

\end{document}